\RequirePackage{fix-cm}
\documentclass[smallextended]{svjour3}       
\smartqed  
\usepackage{graphicx}
\usepackage{amssymb}
\usepackage{latexsym,amsbsy,amsmath,mathtools}
\usepackage{multirow}
\def\vec#1{\boldsymbol{#1}}
\def\llcol#1#2{\tilde{\lambda}_{#1}.\tilde{\lambda}_{#2}}
\journalname{Few Body Syst.}
\begin{document}
\title{Hidden and open heavy-flavor hadronic states}
\author{H.~Garcilazo         \and
A.~Valcarce
}
\authorrunning{H.~Garcilazo {\em et al.}}
\institute{H.~Garcilazo \at
				   \author{H.~Garcilazo} 
           Escuela Superior de F\' \i sica y Matem\'aticas, \\ 
           Instituto Polit\'ecnico Nacional, Edificio 9, 
           07738 M\'exico D.F., Mexico\\ 
           \email{humberto@esfm.ipn.mx}
					 \and
					 A.~Valcarce \at 
           Departamento de F\'\i sica Fundamental,\\ 
           Universidad de Salamanca, E-37008 Salamanca, Spain\\
           \email{valcarce@usal.es}					 
}
\date{Received: date / Accepted: date}
\maketitle
\begin{abstract}
We discuss the stability of hidden and open heavy-flavor hadronic states
made of either two or three mesons. References are made in passing to studies
regarding two and three-body systems containing baryons.
We perform a comparative study analyzing the results in terms of quark and hadron 
degrees of freedom. Compact and molecular states are found to exist in
very specific situations. 
We estimate the decay width for the different scenarios: 
weak decays for bound states by the strong interaction, and strong decays
for hadronic resonances above a decay threshold. The experimental observation of narrow hadrons 
lying well above their lowest decay threshold is theoretically justified.
\end{abstract}

\section{Introduction}
\label{sec0}
The hadron spectra above open-flavor thresholds has emerged as a key issue to
understand QCD in the low-energy regime. The experimental hadron spectra 
below open-flavor thresholds follow closely a naive quark-antiquark ($q\bar q$) 
or three-quark ($qqq$) 
structure according to $SU(3)$ irreducible representations~\cite{Gel64}.
However, since 2003, several resonances reported 
by different experimental collaborations appeared close to a two-hadron threshold,
presenting properties that makes a naive quark substructure unlikely. See, for example, 
Refs.~\cite{Che16,Bri16,Ric16,Leb17,Ali17,Esp17,Liu19} and references therein. Although
this observation could be coincidental due to the large number of open-flavor 
thresholds in the energy region where the new intriguing states have been reported, 
it could also point to a close relation between some particular thresholds and 
resonances contributing to the standard hadron spectroscopy.

The possible existence of hadrons with a quark content richer than $q\bar q$  
or $qqq$ states is nowadays a hot topic in hadron spectroscopy. 
Experimental discoveries have stimulated a flurry of theoretical studies 
dealing with multiquark states and hadron-hadron resonances with a variety
of different methodological approaches.
It is important to note at the outset that conclusions drawn from 
hadron-hadron resonance analyses or a multiquark constituent picture should be similar, 
provided that, in general, a coupled-channel hadron-hadron approach would be 
mandatory in order to reproduce the multiquark constituent picture.
To be more specific, let us note that multiquark systems present a 
richer color structure than standard baryons or mesons. 
Whereas the color wave function for standard mesons and baryons 
is made of a single vector, for multiquark states there are 
different vectors leading to a color singlet.
For example, for four-quark states one can get a color singlet 
out of colorless singlet-singlet ($ 1 1$) or colored ($8 8$, $\bar 3 3$, or $6 \bar 6$) components. 
Any colored component, better known as hidden-color vectors,
can be expanded in terms of colorless singlet-singlet 
states~\cite{Har81,Via09} leading to a coupled-channel
problem at hadronic level. Thus, an important question 
is whether one is in front of a colorless molecule or a 
compact state. Besides the color components, also the 
spatial distribution of the internal quark clusters is 
of great help to discriminate between the two structures~\cite{Via09}. 

Recently, the widely tackled sector of exotic states with two units of 
flavor, $QQ\bar q \bar q$, has been revitalized while 
their non-exotic partners, $Qq\bar Q\bar q$, have 
been much discussed in the context of the $XYZ$ 
mesons~\cite{Che16,Bri16,Ric16,Leb17,Ali17,Esp17,Liu19}. 
In this contribution, we discuss the stability patterns of
hidden and open heavy-flavor hadronic states made of two mesons,
$M_1M_2$ and $M_1\bar M_2$~\cite{Val18}, and 
three mesons, $M_1M_2M_3$~\cite{Gac18}.
References are made in passing to studies
regarding two and three-body systems containing baryons~\cite{Gar84,Caa12,Gar17,Ric17}.
We infer overriding trends which are intended to reflect overall
properties of the systems under study beyond peculiarities of a particular model.
Thus, we will try to link results obtained using quark degrees 
of freedom with those derived in hadronic approaches using the 
common-sense rule of taking the same pairwise interaction.
At present, this connection is a missing link in studies of 
low-energy hadron structure, and should be dealt with 
vigorously from the outset. This work could be a useful 
contribution to allow preliminary conclusions to be drawn. 
Finally, we will also present 
general results for the decay width of bound states by the
strong interaction and hadronic resonances above a 
decay threshold~\cite{Gac17,Gar18,Her20}. 
\section{General rationale based on symmetry breaking}
\label{sec1}
The analogy between the stability of few-charge systems and multiquarks 
in additive spin-independent potentials provides guidance on how to 
identify the favorable multiquark configurations that can lodge 
hadronic resonances and/or bound states. There are, 
however, some differences, not so much due to the radial shape of the 
potential, but mainly due to the color algebra replacing the simpler
algebra of electric charges. The internal dynamics of multiquark states is largely
unknown and thus relies on some extrapolation from models
that correctly accounts for the properties of ordinary
mesons and baryons. The simplest and most widely 
used option consists of two-body potentials with simple 
color dependence, including both a spin-independent (chromoelectric) 
and a spin-dependent (chromomagnetic) component. 
We shall adopt here the so-called AL1 model by Semay and 
Silvestre-Brac~\cite{Sem94}. It includes a standard
Coulomb-plus-linear central potential, supplemented 
by a smeared version of the chromomagnetic interaction,
\begin{eqnarray}
V(r)  & = &  -\frac{3}{16}\, \llcol{i}{j}
\left[\lambda\, r - \frac{\kappa}{r}-\Lambda + \frac{V_{SS}(r)}{m_i \, m_j}  \, \vec \sigma_i \cdot \vec \sigma_j\right] \, , \nonumber\\
V_{SS}(r)  & = & \frac{2 \, \pi\, \kappa^\prime}{3\,\pi^{3/2}\, r_0^3} \,\exp\left(- \frac{r^2}{r_0^2}\right) ~,\quad
 r_0 =  A \left(\frac{2 m_i m_j}{m_i+m_j}\right)^{-B}\!,
\label{ER_1}
 \end{eqnarray}
where $\lambda=$ 0.1653~GeV$^2$, $\Lambda=$ 0.8321~GeV, $\kappa=$ 0.5069, $\kappa^\prime=$ 1.8609,
$A=$ 1.6553~GeV$^{B-1}$, $B=$\,0.2204, $m_u=m_d=$ 0.315~GeV, $m_s=$ 0.577~GeV, $m_c=$ 1.836~GeV 
and $m_b=$ 5.227~GeV. 
Here, $\llcol{i}{j}$ is a color factor, suitably modified for the quark-antiquark pairs.
The smearing parameter of the spin-spin term is adapted to 
the masses involved in the quark-quark or quark-antiquark pairs. 
The parameters of the AL1 potential are constrained in a 
simultaneous fit of 36 well-established mesons 
and 53 baryons, with a remarkable agreement 
with data, see Table 2 of Ref.~\cite{Sem94}.

To compute the ground-state of a $q \bar q$ meson or a $qqq$ baryon 
in a constituent model, a 
crude variational approximation is often sufficient. For systems with a larger number of
constituents the situation is drastically different. For example, 
for a tetraquark close to its threshold, one has to estimate 
precisely $q_1q_2\bar q_3\bar q_4$ and its thresholds, to see whether there is a bound state. 
Moreover,  the $q_1q_2\bar q_3\bar q_4$ wave function has a $(q_1\bar q_3)(q_2\bar q_4)$ 
component and a $(q_1\bar q_4)(q_2\bar q_3)$ one, corresponding to its {\it molecular} part, 
perhaps a $(q_1 q_2)(\bar q_3\bar q_4)$ {\it diquark-antidiquark} component, and a {\it collective} 
component that prevails in the event of deep binding. 

Given that the chromomagnetic forces vanish in the limit of very heavy quarks,
see Eq.~\eqref{ER_1}, it is instructive to consider the case of a purely chromoelectric interaction
to guide the search for optimal configurations to host hadronic resonances.
Under these conditions, firm theoretical conclusions can be obtained.
In quantum mechanics, it is well-known that breaking a symmetry lowers
the ground-state energy~\footnote{For instance, 
going from $H_0=p^2+x^2$ to $H_0 + \lambda x$, lowers the ground-state energy
from $E_0=1$ to $E_0-\lambda^2/4$, and more generally, breaking parity in 
$H=H_{\rm even} + H_{\rm odd}$ gives $E < E_{\rm even}$.}. But in a few-body system,
the breaking of symmetry often benefits more to the threshold than to the collective 
configuration and thus spoils the binding. From these results, one can analyze the effect 
of symmetry breaking in systems of 
four-charged particles. Let us first consider the hydrogen molecule, $M^+ M^+ m^- m^-$.
The Hamiltonian for this system reads,
\begin{eqnarray}
H&=&\frac{\vec p_1^{\,2}}{2 \, M} +
\frac{\vec p_2^{\,2}}{2 \, M} +
\frac{\vec p_3^{\,2}}{2 \, m} +
\frac{\vec p_4^{\,2}}{2 \, m} + V = H_0 + H_1 \nonumber \\
&=&\left[\sum_{i}\frac{\vec p_i^{\, 2}}{2 \, \mu} \, + \, V \right] \, + \, \left(\frac{1}{4\,M} \, - \, \frac{1}{4\,m} \right)\left(
\vec p_1^{\,2} + \vec p_2^{\,2} - \vec p_3^{\,2} - \vec p_4^{\,2} \right) \, ,
\label{ERN_1}
\end{eqnarray}
where $2\,\mu^{-1}=M^{-1}+m^{-1}$. The $C$-parity breaking term, $H_1$, lowers the ground-state
energy of $H$ with respect to the $C$-parity even part, $H_0$, which is simply a rescaled version of
the Hamiltonian of the positronium molecule. Since $H_0$ and $H$
have the same threshold, and since the positronium molecule is stable, the hydrogen molecule
is even more stable, and stability improves when $M/m$ increases. 
Clearly, the Coulomb character of $V$ hardly matters in this reasoning.  
The key property is that the potential does not change when the masses are modified,
a property named {\it flavor independence} in QCD.

One can use the same reasoning to study the stability of four-charged particles 
when $C$-parity is preserved but particle symmetry is broken, in other words the
$M^+ m^+ M^- m^-$ configuration. The Hamiltonian is given by,
\begin{eqnarray}
H&=&\frac{\vec p_1^{\,2}}{2 \, M} +
\frac{\vec p_2^{\,2}}{2 \, m} +
\frac{\vec p_3^{\,2}}{2 \, M} +
\frac{\vec p_4^{\,2}}{2 \, m} + V = H_0 + H_1 \nonumber \\ 
&=&\left[\sum_{i}\frac{\vec p_i^{\, 2}}{2 \, \mu} \, + \, V \right] \, + \, \left(\frac{1}{4\,M} \, - \, \frac{1}{4\,m} \right)\left(
\vec p_1^{\,2} + \vec p_3^{\,2} - \vec p_2^{\,2} - \vec p_4^{\,2} \right) \, .
\label{ERN_2}
\end{eqnarray}
On the basis of the arguments made above it is right to conclude that the 
ground-state of $H$ gains binding with respect to the threshold 
$(M^+m^-) - (M^-m^+)$ that it shares with $H_0$.
However, there is another threshold that lies lower, $(M^+M^-) - (m^+m^-)$. This 
threshold gains more from the symmetry breaking than the four-body molecule, and, 
indeed, it is found that the molecule becomes unstable for $M/m \gtrsim 2.2$. In other 
words, a protonium atom cannot polarize enough
a positronium atom and stick to it. It remains that the 
hydrogen-antihydrogen system could form a kind of metastable molecule 
below the atom-antiatom threshold~\cite{Val18,Gar18}. 

In a semirelativistic framework the decomposition of $H$ into a 
symmetric and an antisymmetric part under $C$-parity still holds, 
and the antisymmetric part lowers the ground state energy. However,
$H$ has not the same threshold as its symmetric part and
one should study what wins, the asymmetry in the 4-body Hamiltonian 
or the one in the 2-body Hamiltonian~\cite{Ric20}.

\section{Two-meson states}
\label{sec2}

\subsection{Two-meson compact states: $QQ\bar q\bar q \equiv M_1M_2$}
\label{sec2-1}

The arguments set out above after Eq.~\eqref{ERN_1}, can be directly 
translated to four-quark systems:
the $QQ\bar q \bar q$ configuration  becomes more and more
bound when the mass ratio $M/m$ increases. This has been established 
in the pioneering work of Ref.~\cite{Ade82}, and 
discussed and confirmed in further studies~\cite{Ric18}\footnote{It is worth to 
note that Ref.~\cite{Gar84} derived the same conclusion for a three-body
system of particles with masses $MM\mu$, with $M > \mu$. For non-interacting 
heavy-particles and an slightly attractive 
mass independent interaction between the light and heavy particles, 
the binding energy of the three-body system
increases rapidly when $M/\mu$ augments, see Fig. 3 of Ref.~\cite{Gar84}.}. 
A remaining problem would be to understand why the positronium
molecule lies slightly below its dissociation threshold, while a chromoelectric
model associated with the color additive rule does not bind (at least according 
to most computations~\cite{Ric18}). This is due to a 
larger disorder in the color coefficients than in the electrostatic
strength factors entering the Coulomb potential~\cite{Val18}. 
Thus, multiquarks are penalized by the non-Abelian character of the color algebra, 
and its stability cannot rely on the asymmetries of the potential energy. 
It should use other asymmetries, in particular through the masses entering the kinetic 
energy, chromomagnetic effects generating mixing of $\bar 3 3$ and $6 \bar 6$
or the coupling to decay channels, etc.

\begin{table}[t]
\caption{Properties of the $QQ\bar q \bar q$ ground state as a function of the mass of the
heavy quark $M_Q$ for the AL1 model~\cite{Sem94}. Energies and masses are in MeV and distances in fm.}
\begin{center}
	\renewcommand{\arraystretch}{1.05}
  \begin{tabular}{ccccccccccc}
\hline
$M_Q$  &
$Th$\hspace*{-5pt} &
\multicolumn{1}{c}{$\Delta E$}    &
$P[| \bar 3 3\rangle]$ & $P[| 6 \bar 6\rangle]$ &
$P_{MM^*}$ & $P_{M^*M^*}$ &
$\bar x$ & $\bar y$ & $\bar z$ 
\\ \hline
5227	& 10644	& $-$151 & 0.967 &	0.033 &	0.561  &	0.439 &	0.334 & 0.784 &	0.544 \\ 
4549	&	9290  & $-$126 & 0.955 &	0.045 &	0.597  &	0.403 &	0.362 &	0.791 & 0.544 \\
3871	&	7936  & $-$100 & 0.930 &	0.070 &	0.646  &	0.354 &	0.411 &	0.806 & 0.541 \\
3193	&	6582	& $-$71	 & 0.885 &	0.115 &	0.730  &	0.270 &	0.475 &	0.833 & 0.536 \\
2515	&	5230	& $-$41	 & 0.778 &	0.222 &	0.795  &	0.205 &	0.621 &	0.919 & 0.523 \\
1836	&	3878	& $-$13	 & 0.579 &	0.421 &	0.880  &  0.120 &	0.966 &	1.181 & 0.499 \\
1158	&	2534	& $>$ 0  & 0.333 &	0.667 &	1.000  &	0.000 &$\gg$ 1&$\gg$ 1& 0.470 \\ \hline
\end{tabular}
\end{center}
\label{TR_1}
\end{table}

We present in Table~\ref{TR_1} the results for the ground state of a $QQ\bar q \bar q$ system
with the AL1 potential~\cite{Sem94}, an isoscalar $J^P=1^+$ state, as a function of the mass 
of the heavy quark $M_Q$~\cite{Her20}. For each value 
of $M_Q$ we have evaluated the lowest
strong-decay threshold, $Th= M_1 + M_2$, and the corresponding binding energy $B=-\Delta E$. 
The binding energy increases with increasing $M_Q/m_q$ ($m_q$ is kept constant). 
Close to $\Delta E = 0$ the system behaves like
a simple meson-meson molecule, with a large probability in a single meson-meson
component, the pseudoscalar-vector channel $MM^*$. However,
when $M_Q/m_q$ increases the probability of the $6 \bar 6$ color component
diminishes (it tends to zero for $M_Q \to \infty$). Therefore, heavy-light compact bound 
states would be almost a pure $\bar 3 3$ color state and not a 
single colorless meson-meson molecule, $ 1 1$.
Such compact states with internal colored components can be expanded in terms of
physical meson-meson channels~\cite{Har81,Via09}, in this case pseudoscalar-vector $MM^*$ and
vector-vector $M^*M^*$ components. Thus, $QQ\bar q \bar q$ states in the
limit of large binding, i.e., $M_Q/m_q$ large, can also be 
studied as a coupled-channel problem of physical meson-meson states
leading to the same results~\cite{Car11,Vij14,Ike14}. Note however
that the interaction between the clusters should be derived from the
basic interactions between the constituents, which must be submitted to
antisymmetrization~\cite{Car11}. Thus, a narrow spatial distribution 
would be also obtained~\cite{Via09}.   

\begin{figure}[t]
\centering
\vspace*{-.75cm} 
\includegraphics[width=0.95\columnwidth]{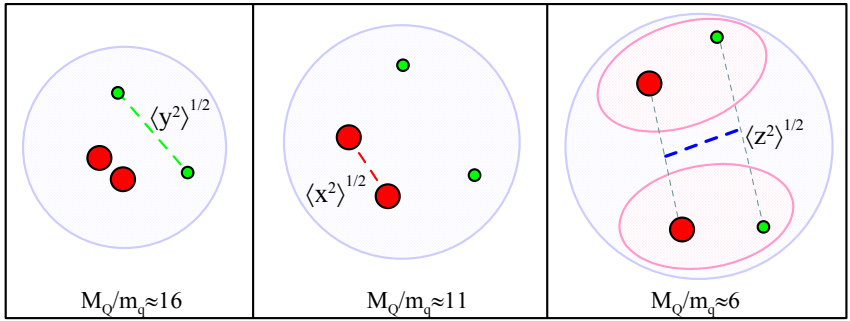}
\vspace*{-10.75cm}
\caption{Schematic representation of the internal structure of the $QQ\bar q \bar q$ 
ground state as the heavy-quark mass decreases and, accordingly, the binding energy.}
\label{FR_1}
\end{figure}

The results in Table~\ref{TR_1} show how when the binding increases, i.e. $M_Q/m_q$
augments, the average distance between the two heavy quarks, $\bar x = \langle x^2\rangle^{1/2}$, diminishes
rapidly, while that of the two light quarks, $\bar y = \langle y^2\rangle^{1/2}$, although diminishing,
remains larger. The heavy-to-light quark distance, $\bar z = \langle z^2\rangle^{1/2}$, stays almost 
constant for any value of $M_Q/m_q$. 
Thus, in the heavy-quark limit, the lowest lying tetraquark configuration
resembles the Helium atom~\cite{Lip86,Eic17,Qui18}, a factorized system with separate dynamics
for the compact color $\bar 3$ $QQ$ {\it kernel} and for the
light quarks bound to the stationary color $3$ state, to construct
a $QQ\bar q \bar q$ color singlet. These results
present a sharp picture of how the internal structure of the $QQ\bar q\bar q$ 
ground state changes according to the ratio $M_Q/m_q$,
in other words, from a deeply bound compact state to a 
close-to-threshold meson-meson molecule, see Fig.~\ref{FR_1}. 

A similar situation appears in the case of non-identical heavy-flavor mesons, or conversely
systems of the type $QQ'\bar q \bar q$~\cite{Car19}. Due to the existence 
of two distinguishable heavy quarks, there appears a strong-interaction
stable $J^P=0^+$ state below the $1^+$ state discussed above for identical heavy quarks. 
The larger number of 
basis vectors contributing to a particular set of quantum numbers, some of which are 
forbidden in a system with identical heavy flavors, is relevant to understand
how $QQ'\bar q\bar q$ bound states are formed. 
\begin{figure}[t]
\centering
\vspace*{-2.9cm}
\includegraphics[width=0.8\columnwidth]{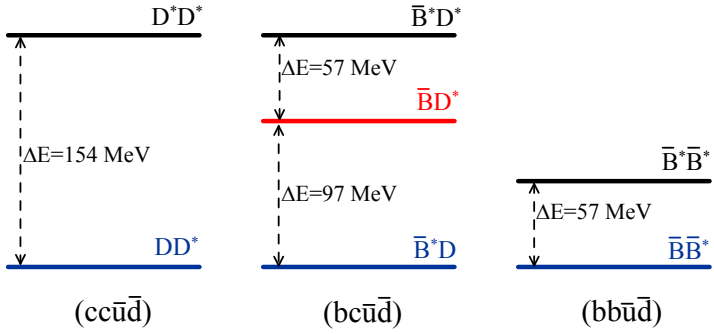}
\vspace*{-6.0cm}
\caption{Two-meson thresholds for the isoscalar $J^P=1^+$ 
$cc\bar u\bar d$, $bc\bar u\bar d$, and $bb\bar u\bar d$ states.}
\label{FR_2}
\end{figure}

Let us analyze how the dynamics of thresholds, see Fig.~\ref{FR_2},
helps to understand the results obtained for the $QQ'\bar q\bar q$ system.
For this purpose we focus on the isoscalar $J^P=1^+$
bound state, that exists both with identical ($bb$ and $cc$) and non-identical ($bc$)
heavy flavors. 
The chromomagnetic interaction, suppressed by $M_Q$ as seen in Eq.~\eqref{ER_1}, 
generates larger matrix elements in the charm than in the bottom sector
between color-spin vectors of the pseudoscalar-vector and vector-vector 
two-meson components. However, as the mass difference between the two-meson 
components increases from 57~MeV in the bottom sector to 154~MeV in the 
charm one~\footnote{Results obtained with
the AL1 model~\cite{Sem94}.}, the coupling effect is weakened~\cite{Car16}. 
Since the single channel problem of $D D^*$
or $\bar B \bar B^*$ mesons does not present bound states~\cite{Via09,Car11},
the weaker chromomagnetic coupling between $D D^*$ and $D^* D^*$ 
than between $\bar B\bar B^*$ and $\bar B^*\bar B^*$, leads to
a reduction of the binding energy from 151~MeV in the bottom sector 
to 13~MeV in the charm one, see Table~\ref{TR_1}. 

If we now 
consider the isoscalar $bc\bar u\bar d$ $J^P=1^+$
state, the mass difference between $\bar B^* D$ and $\bar B^* D^*$ 
is the same as in the charm case, but
the chromomagnetic interaction involving the bottom quark 
is weakened by a factor $m_b/m_c \sim 3$. Thus,
a smaller binding energy than in the charm sector would be expected. However, the results
exhibit a different trend, with a larger binding energy of 23~MeV~\cite{Car19}. 
What it is different about the $bc\bar q\bar q$ system is that it contains distinguishable 
heavy quarks and thus a new pseudoscalar-vector two-meson component (note
that $\bar M_1 M_2^*$ and $\bar M_1^* M_2$ have now a different mass) 
contributes to the $J^P=1^+$ state. Besides, this new two-meson component, 
the $\bar B D^*$, is in between $\bar B^* D$ and $\bar B^* D^*$, see Fig.~\ref{FR_2}.
Interestingly enough, 
although the $\bar B^* D$ and $\bar B D^*$ states are not 
directly coupled, nevertheless, they become indirectly coupled 
through the higher $\bar B^* D^*$ state, i.e. $\bar B^* D\leftrightarrow \bar B^* D^* \leftrightarrow \bar B D^*$.
Being the mass difference between $\bar B^* D$ and $\bar B D^*$ smaller than between $D D^*$ and $D^* D^*$
the mixing is reinforced as compared to the charm case, leading to a binding energy larger than in the 
charm sector. The dynamics of thresholds to enhance or diminish coupled-channel effects has been
illustrated at length in the literature~\cite{Gar17,Lut05,Gar07,Gac07,Gar15,Bar15,Car16}.
However, Ref.~\cite{Car19} reported the first example where the presence of an additional 
intermediate threshold induced by the non-identity of the heavy quarks
helps increasing the binding.
\begin{figure}[t] 
\vspace*{-2.6cm}
\hspace*{-0.8cm}\resizebox{6.5cm}{9.3cm}{\includegraphics{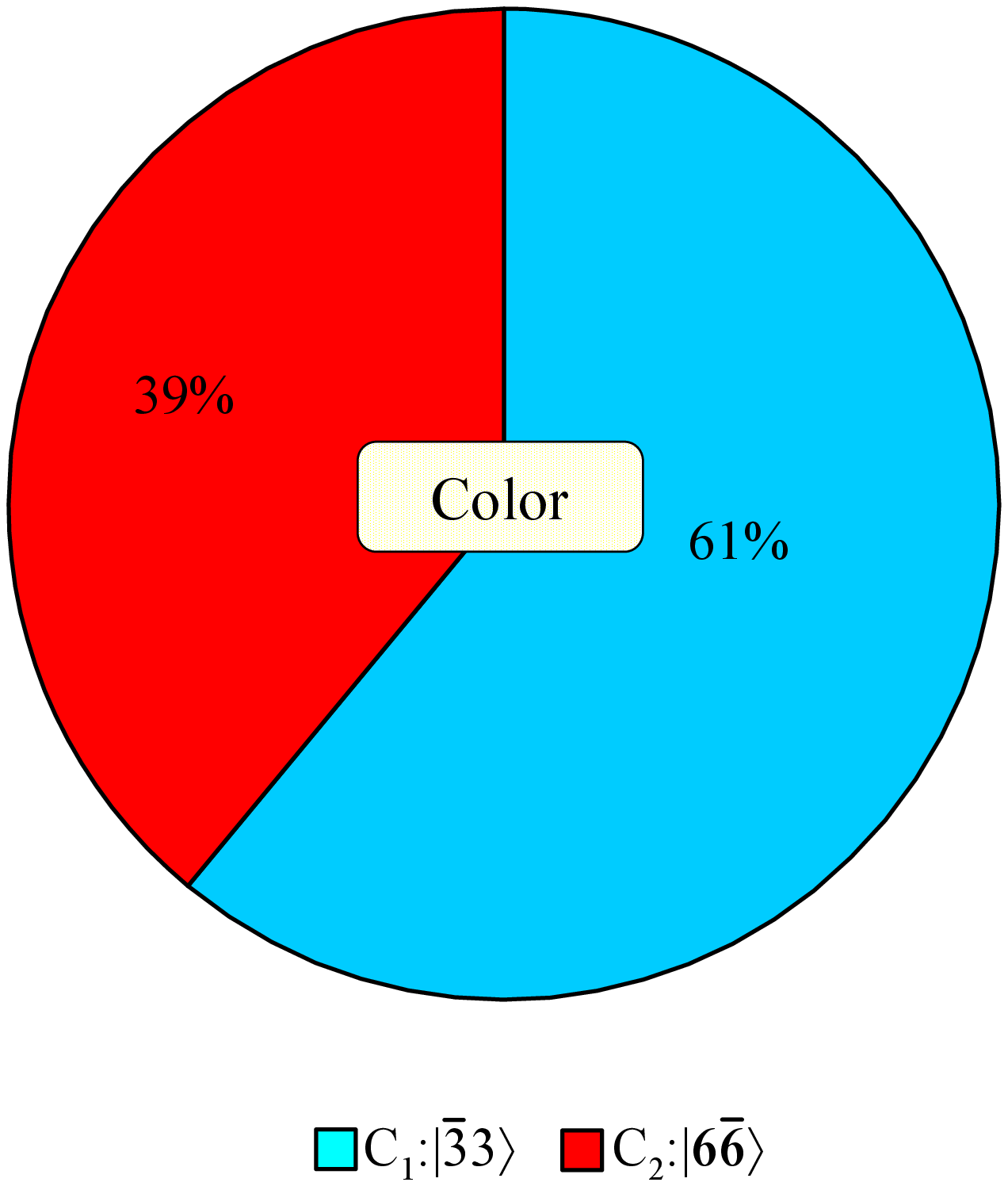}}
\resizebox{6.5cm}{9.3cm}{\includegraphics{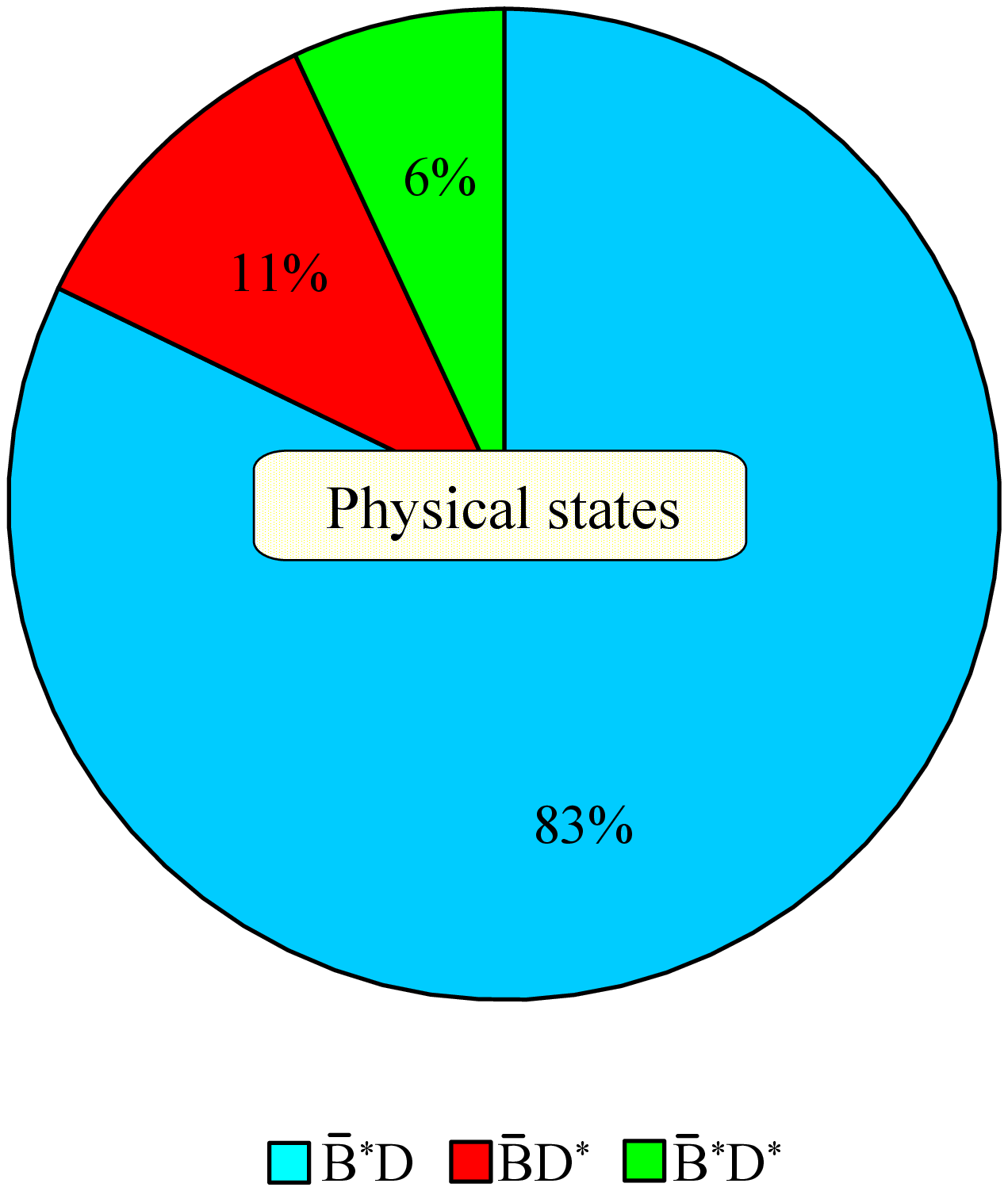}}
\vspace*{-2.0cm}
\caption{Detailed structure of the isoscalar $bc\bar u\bar d$ $J^P=1^+$ color wave function,
showing the decomposition in terms of singlet-singlet color vectors, i.e., 
physical states~\cite{Via09}.}
\label{FR_3}
\end{figure}

Thus, there is an obvious link between studies based on
quark degrees of freedom and those relying on hadronic models 
provided a coupled-channel approach is followed in the hadronic description.
The equivalence can be analytically derived through the formalism 
developed in Ref.~\cite{Via09}. It allows 
to extract the probabilities of meson-meson physical channels out of a four-quark wave function 
expressed as a linear combination of color-spin-flavor-radial vectors. We show in Fig.~\ref{FR_3} a 
summary of the color and meson-meson 
component probabilities for the isoscalar $J^P=1^+$ $bc\bar u\bar d$ bound state. 
It is worth noting the 11\% probability of the $\bar B D^*$ component, induced by the indirect coupling
to the lowest $\bar B^* D$ state through the highest $\bar B^* D^*$ component.
As has been recently discussed~\cite{Ric18}, these results present sound evidence
about the importance of including a complete basis, i.e., not discarding any set of 
basis vectors a priori. Unless it is done that way, one is in front of approximations 
driving to unchecked results.

\subsection{Two-meson molecular states: $Qq\bar Q\bar q \equiv M\bar M$}
\label{sec2-2}
\begin{figure*}[t]
\centering
\includegraphics[width=12cm,height=15cm]{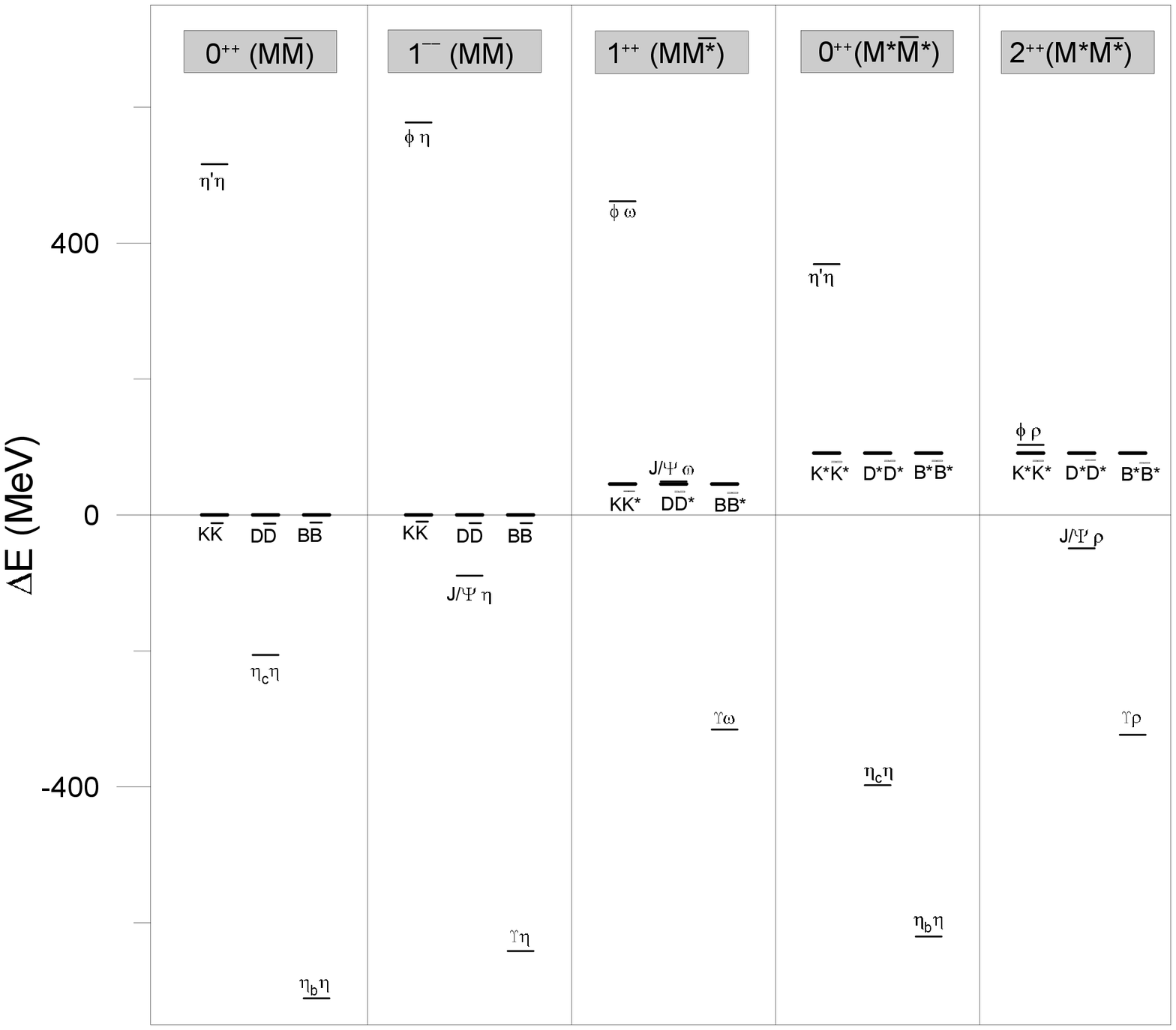}
\vspace*{-6.0cm}
\caption{Experimental masses~\cite{Tan18} of the different two-meson $Qq\bar Q \bar q$ systems with 
$Q=s$, $c$, or $b$, for several sets of quantum numbers, $J^{PC}$. The 
reference energy has been set to the $K\bar K$, $D\bar D$ and $B\bar B$ masses for the hidden
strange, charm and bottom sectors, respectively.}
\label{FR_4}
\end{figure*}

As discussed in Sect.~\ref{sec1} after Eq.~(\ref{ERN_2}), the presence of 
two thresholds for $Qq\bar Q\bar q$, 
one of them taking benefit from
the breaking of the particle identity, makes a priori the stability of meson-antimeson 
states much more difficult. Hadrons with a $Qq\bar Q \bar q$ 
flavor content, could split either into $(Q \bar q) - (q \bar Q)$ or
$(Q \bar Q) - (q \bar q)$ two-meson states~\cite{Car12}. For $Q=c$, the $(Q \bar Q) - (q \bar q)$ 
and $(Q \bar q) - (q \bar Q)$ thresholds are almost degenerate, 
while for $Q=b$ the $(Q \bar Q) - (q \bar q)$ 
threshold is much lower than the $(Q \bar q) - (q \bar Q)$ 
one as shown in Sect.~\ref{sec1}. See also Fig.~\ref{FR_4}.
The growth of the mass difference between the two thresholds
when the mass of the heavy quark increases
is linked to the flavor-independence of the 
chromoelectric interaction~\cite{Eic75,Isg99,Ber79,Nus99}. 
Thus, the possibility of finding stable 
meson-antimeson molecules, $(Q \bar q) - (q \bar Q)$, becomes more 
difficult when increasing the mass of the heavy flavor unless the two thresholds
$(Q \bar q) - (q \bar Q)$ and $(Q \bar Q) - (q\bar q)$ would be decoupled and 
thus a narrow quasibound state may arise~\cite{Gar18}, as 
will be discussed in Section~\ref{sec5}.
\begin{figure*}[t]
\vspace*{-1cm}
\resizebox{5.7cm}{7.5cm}{\includegraphics{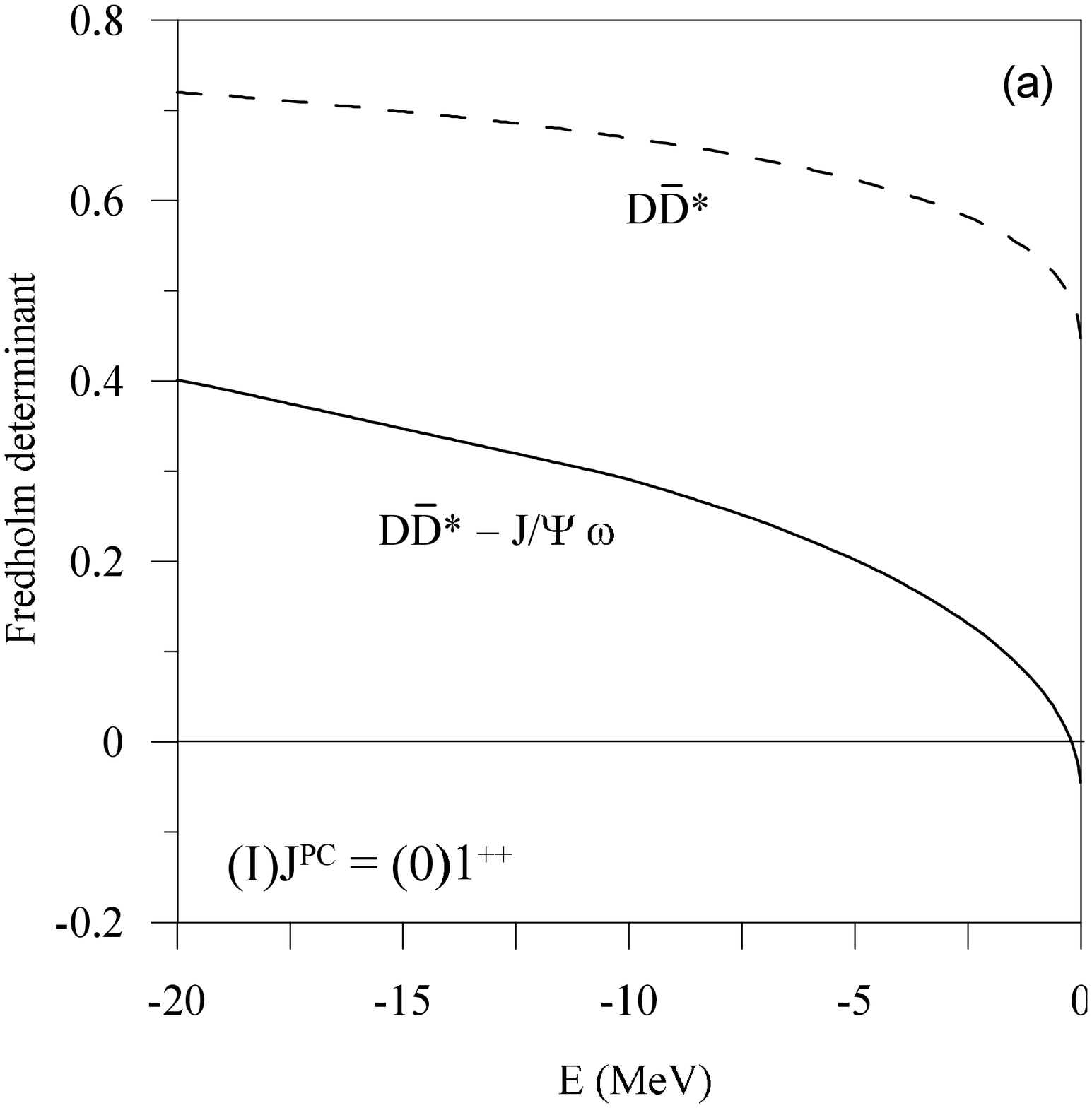}}\qquad
\resizebox{5.7cm}{7.5cm}{\includegraphics{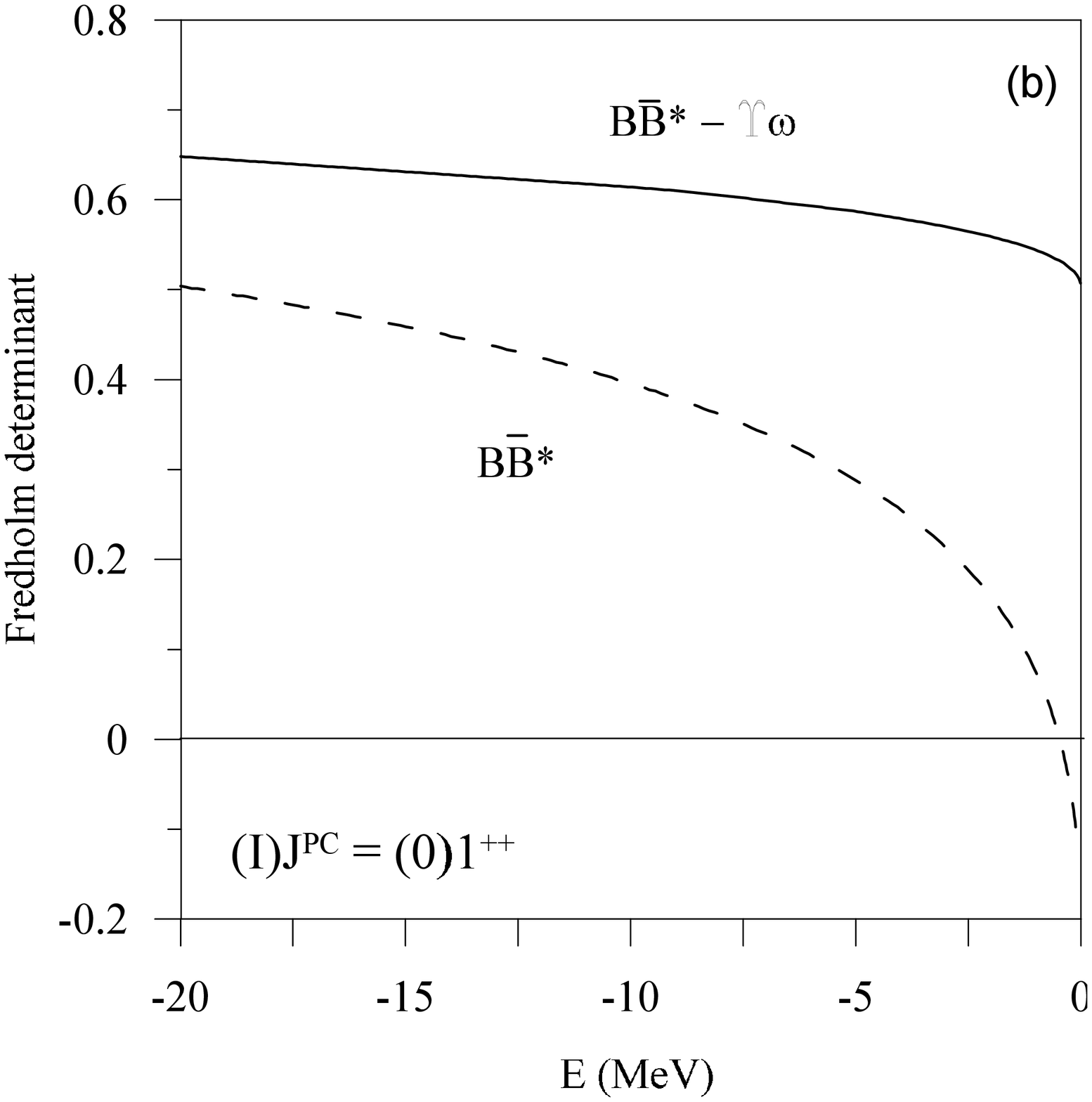}}
\vspace*{-1.5cm}
\caption{(a) $(I)J^{PC}=(0)1^{++}$ $cq\bar c\bar q$ Fredholm determinant. The dashed line 
stands for a calculation considering only charmed mesons, $D\bar D^*$, whereas the solid line
includes the coupling to the $J/\Psi \,\, \omega$ two-meson state. (b) Same
as (a) for bottomonium.}
\label{FR_5}
\end{figure*}

The experimental scenario illustrated in Fig.~\ref{FR_4} 
suggests different consequences for meson-antimeson molecules.
First, the possible existence of stable molecules in the 
hidden-strange sector. If the $K\bar K$ interaction were attractive, 
this two-meson system may be stable because 
no any other threshold appears below.
This was precisely the idea suggested by Weinstein and
Isgur~\cite{Wei90} as a plausible explanation of the proliferation of 
scalar mesons in the light sector. 
Second, the possibility of finding meson-meson molecules contributing to 
the charmonium spectrum due to the coupled-channel dynamics, as in the case
of the $(I)J^{PC}=(0)1^{++}$ quantum numbers. In this case the isoscalar 
$(c \bar q) - (q\bar c) \equiv D\bar D^*$ and $(c\bar c) - (q\bar q) \equiv J/\Psi\,\, \omega$ 
two-body channels are so close together~\footnote{Note that the vicinity of
thresholds is a necessary though not sufficient condition for the existence of a 
resonance. See Ref.~\cite{Vij14} for a thorough and critical analysis.} 
that a slightly attractive interaction along
with the cooperative effect of the almost degenerate two-body channels
provides a plausible explanation of the $X(3872)$~\cite{Bra08,Car09}. 

In spite of the widespread belief that the stability of a multiquark state
is favored by increasing the mass of the heavy flavors, the structures studied in
this section send a clear warning that it is not always the case~\footnote{See, for example,
Ref.~\cite{Rih17} for a further demonstration of the instability of all-heavy tetraquarks
$QQ\bar Q \bar Q$ with a rigorous treatment of the few-body problem.}.
The reason is that the mass of one of the thresholds,
$(Q\bar Q) - (q\bar q)$, diminishes rapidly when the heavy quark mass increases.
This simple reasoning, formulated in terms of coupled-channel arguments,
is illustrated in Fig.~\ref{FR_5}. In Fig.~\ref{FR_5}(a) it can be seen how the 
$D\bar D^*$ interaction (dashed line) is not attractive 
enough to generate a bound state (the Fredholm determinant falls short of 
being negative~\cite{Gar87}). The coupling to 
the $J/\Psi \, \omega$ channel (solid line) 
is responsible for having a bound state just
below threshold. Note that this plausible explanation of the $X(3872)$ is strengthened
by the subsequent experimental observation of the decay $X(3872) \to J/\Psi \, \omega$~\cite{Amo10}.
When the mass of the heavy quark augments from charm to bottom, 
the $B \bar B^*$ interaction becomes more
attractive, dashed line in Fig.~\ref{FR_5}(b). However, the coupling to the lower channel,
$\Upsilon \omega$, would destroy the possibility of having a bound state, solid line in
Fig.~\ref{FR_5}(b). Thus, based on constituent model arguments, one should 
not expect a twin of the $X(3872)$ in the bottom sector, as pointed out by 
hadronic models based on the traditional meson theory of the nuclear 
forces~\cite{Tor92}  
or resorting to heavy quark symmetry arguments~\cite{Sun11,Nie11}.
Note however, that if the $B\bar B^*$ and $\Upsilon \omega$ channels were 
decoupled, as it is suggested by the rescaling~\cite{Pol18} of recent 
lattice QCD calculations of the interaction of the $J/\Psi$ with nuclear 
matter~\cite{Suf18}, the arguments put forward by Ref.~\cite{Gar18} 
on the basis of the results shown in Fig.~\ref{FR_5}(b) 
would justify that a narrow resonance might appear just below 
the $B\bar B^*$ threshold.

\section{Three-meson bound states}
\label{sec3}
The broad theoretical 
consensus~\cite{Ade82,Ric18,Eic17,Vij09,Fra17,Kar17,Bic16,Luo17,Duc13,Cza18,Jun19}
on the existence of an isoscalar doubly bottom tetraquark, $T_{bb}$~\footnote{The 
binding energy reported for the $T_{bb}$ tetraquark
ranges between 90 and 214 MeV.},
discussed in Sect.~\ref{sec2-1}, opens the door to the possible existence 
of other bound states with a 
larger number of hadrons~\cite{Gac18,Gar17,Mam19}. The answer is by no means trivial. 
Ref.~\cite{Gar17} studied three-body systems containing $D$ and $\bar B$ mesons 
together with nucleons and $\Delta$'s. It was shown that 
if the different two-body thresholds of a three-body system are far away, 
they conspire against the stability of the three-body system. Thus,
in this section we review the stability of systems made of three $B$ mesons.
\begin{table}[t]
\caption{Different two-body channels $(i,j)$ contributing to the $(I)J^P=(1/2)2^-$
$B B^* B^* - B^* B^* B^*$ system.}
\begin{center}
\begin{tabular}{ccc} 
\hline
Interacting pair & $(i,j)$ & Spectator  \\ 
\hline
\multirow{2}{*}{$B B^*$}   & $(0,1)$   & \multirow{2}{*}{$B^*$} \\
                           & $(1,1)$   &                      \\
\multirow{2}{*}{$B^* B^*$} & $(0,1)$   & \multirow{2}{*}{$B^*$} \\
                           & $(1,2)$   &                      \\
$B^* B^*$                  & $(1,2)$   & $B$ \\ \hline																																					
\end{tabular}
\end{center}
\label{TR_2}
\end{table}

We solve exactly the Faddeev equations for the three-meson bound 
state problem~\cite{Gac18} using as input the two-body $t-$matrices of a constituent 
model~\cite{Gar17}.
We select those $(I)J^P$ three-body channels that contain the
$T_{bb}$ state and where two-body subsystems containing two $B$-mesons 
are not allowed, because the $BB$ interaction does not show an attractive character. 
The three-body channel $(I)J^P=(1/2)2^-$ is the only one bringing together all these 
conditions to maximize the possible binding of the three-body system\footnote{Note
that the three-body channels with $J=0$ or $1$ would couple to two $B$-meson subsystems where 
no attraction has been reported~\cite{Ade82,Ric18,Eic17,Vij09,Fra17,Kar17,Bic16,Luo17,Duc13,Cza18,Jun19},
whereas the $J=3$ would not contain a two-body subsystem with $j=1$, the quantum numbers
of the $T_{bb}$ tetraquark. The same reasoning excludes the $I=3/2$ channels.}. 
We indicate in Table~\ref{TR_2} the two-body channels contributing to this state 
that we examine in the following, leading to a coupled-channel
problem of pseudoscalar-vector and vector-vector 
two $B$-meson components.

\begin{figure}[t]
\centering
\includegraphics[width=.85\columnwidth]{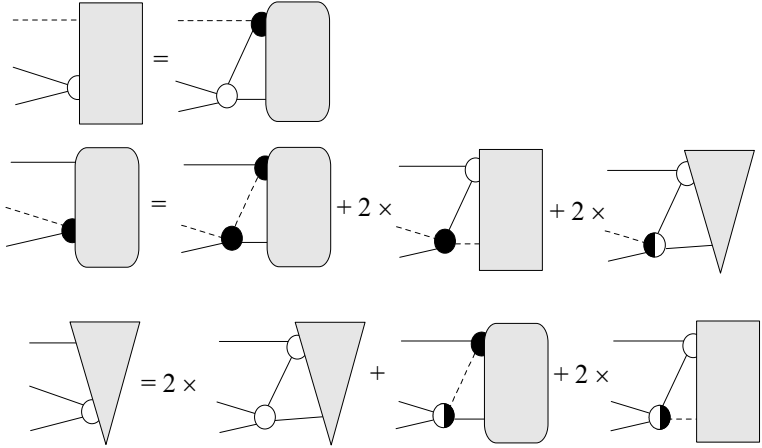}
\vspace*{-8.0cm}
\caption{Diagrammatic Faddeev equations for the three $B$-meson system.}
\label{FR_6}
\end{figure}

The Lippmann-Schwinger equation for the bound-state three-body problem is
\begin{equation}
T=(V_1+V_2+V_3)G_0 T \, ,
\label{ER_2}
\end{equation}
where $V_i$ is the potential between particles $j$ and $k$ and $G_0$ is the
propagator of three free particles. The Faddeev decomposition of
Eq.~\eqref{ER_2},
\begin{equation}
T=T_1 + T_2 + T_3 \, ,   
\label{ER_3}
\end{equation}
leads to the set of coupled equations,
\begin{equation}
T_i = V_i G_0 T\, .
\label{ER_4}
\end{equation}
The Faddeev decomposition guarantees the uniqueness of the solution~\cite{Fad61,Fad65}.
Eqs.~\eqref{ER_4} can be rewritten in the Faddeev form
\begin{equation}
T_i = t_i G _0(T_j + T_k) \, ,   
\label{ER_5}
\end{equation}
with
\begin{equation}
t_i = V_i + V_i G_0 t_i \, ,
\label{ER_6}
\end{equation}
where $t_i$ are the two-body $t-$matrices that already contain the coupling among 
all two-body channels contributing to a given three-body state.
The two sets of equations~\eqref{ER_4} and~\eqref{ER_5} are completely equivalent for the
bound-state problem. In the case of two three-body systems that are coupled together, like $BB^*B^* - B^*B^*B^*$,
the amplitudes $T_i$  become two-component
vectors and the operators $V _i$, $t _i$, and $G _0$ become $2 \times 2$ matrices
and lead to the equations depicted in Fig.~\ref{FR_6}.
The solid lines represent the $B^*$
mesons and the dashed lines the $B$ meson. 
If in the second equation
depicted in Fig.~\ref{FR_6} one drops the last term in the r.h.s. then the first
and second equations become the Faddeev equations of two identical
bosons plus a third one that is different~\cite{Gar17}. Similarly, if 
in the third equation depicted in Fig.~\ref{FR_6} one drops the last two terms
this equation becomes the Faddeev equation of a system of three identical
bosons since in this case the three coupled Faddeev equations are
identical~\cite{Gar17}. The additional terms in Fig.~\ref{FR_6} are, of course,
those responsible for the coupling between the 
$BB^*B^*$ and $B^*B^*B^*$ components. 
\begin{figure}[t]
\centering
\vspace*{-0.3cm}
\includegraphics[width=0.85\columnwidth]{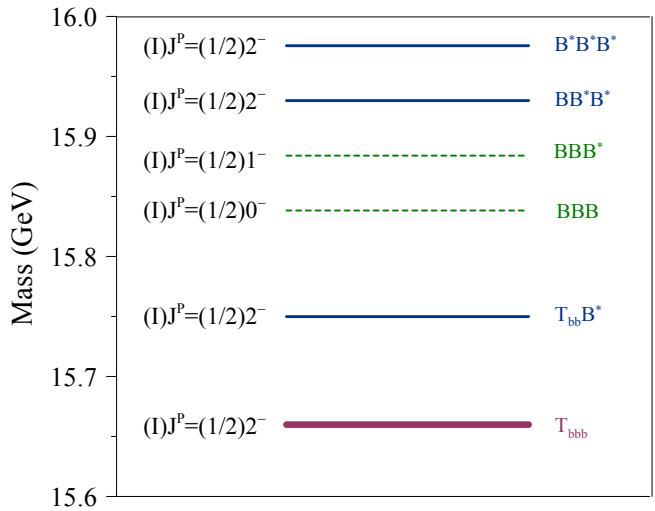}
\vspace*{-7.cm}
\caption{Mass of the three-body $BB^*B^*-B^*B^*B^*$ bound-state $(I)J^P=(1/2)2^-$ 
$T_{bbb}$ (purple thick line),
compared to the different three $B$-meson strong (blue solid lines) 
and electromagnetic (green dashed lines) decay thresholds.}
\label{FR_7}
\end{figure}

We show in Fig.~\ref{FR_7} the results of our calculation. The blue solid
lines stand for the different three $B$-meson strong decay thresholds of the
$BB^*B^*-B^*B^*B^*$ system with quantum numbers $(I)J^P=(1/2)2^-$, that we have denoted by $T_{bbb}$.
These thresholds are $B^*B^*B^*$, $BB^*B^*$ and $T_{bb}B^*$. The green dashed lines stand for the possible 
three $B$-meson electromagnetic
decay thresholds, $BBB^*$ and $BBB$ with quantum number $(I)J^P=(1/2)1^-$ and $(I)J^P=(1/2)0^-$, respectively. 
Finally, the purple thick line indicates the energy of the $T_{bbb}$,
that appears 90~MeV below the lowest threshold. The results shown in Fig.~\ref{FR_7}
correspond to the binding energy of the $T_{bb}$ obtained in Ref.~\cite{Fra17}.
\begin{table}[b]
\caption{Binding energy, in MeV, of the $T_{bbb}$ $(I)J^P=(1/2)2^-$
$B B^* B^* - B^* B^* B^*$ three-body system as a function
of the binding energy, in MeV, of the $T_{bb}$ tetraquark.
The $T_{bbb}$ binding energy is calculated with respect to the 
lowest strong decay threshold: $m_B + 2 \, m_{B^*} - B(T_{bb})$.}
\begin{center}
\begin{tabular}{ccc} 
\hline
$B(T_{bb})$ & $B(T_{bbb})$  \\ 
\hline
180 & 90  \\
144 & 77  \\ 
117 & 57  \\
87  & 43  \\\hline																																					
\end{tabular}
\end{center}
\label{TR_6}
\end{table}

We have checked that the $T_{bbb}$ remains stable for the whole 
range of binding energies of the $T_{bb}$ reported in the literature,
repeating the coupled-channel three-body calculation starting
from the smallest binding of the order of 90~MeV obtained in Ref.~\cite{Bic16}. 
The results are given in Table~\ref{TR_6}. It can be seen how the three-meson 
bound state $T_{bbb}$ is comfortably stable. If the binding 
energy of the $T_{bb}$ is reduced up to 50~MeV, the three-body system
would have a binding of the order of 23~MeV that would already lie 
19~MeV above the lowest $BBB$ threshold, so that one does not expect
any kind of Borromean binding. It is worth noting that many-body 
interactions inspired by the strong-coupling regime of QCD do not
support stability of four-quark exotic states, see the solid line in
Fig. 2 of Ref.~\cite{Vij07} and, thus, they are not 
expected to play a relevant role for the $T_{bbb}$.

If the $T_{bbb}$ would have appeared in between the
two three-body thresholds, $BB^*B^*$ and $B^*B^*B^*$,
it could still be narrow. Ref.~\cite{Gac17} has presented a plausible 
argument based on first-order perturbation theory explaining the 
small width of a three-body resonance in a 
coupled two-channel system lying close to the upper channel in spite 
of being open the lower one. This is a challenging result when the 
available phase space of the decay channel is quite large. 
Similar arguments could be handled for a comprehensive study of the properties
of the LHCb pentaquarks~\cite{Che16,Bri16,Ric16,Leb17,Ali17,Esp17,Liu19,Ric17}.
However, this would require a thorough analysis within each particular model 
used to study these states.

\section{Decay width of $QQ\bar q\bar q$ states}
\label{sec4}
\begin{figure}[b]
 \centerline{
 \includegraphics[width=.47\textwidth]{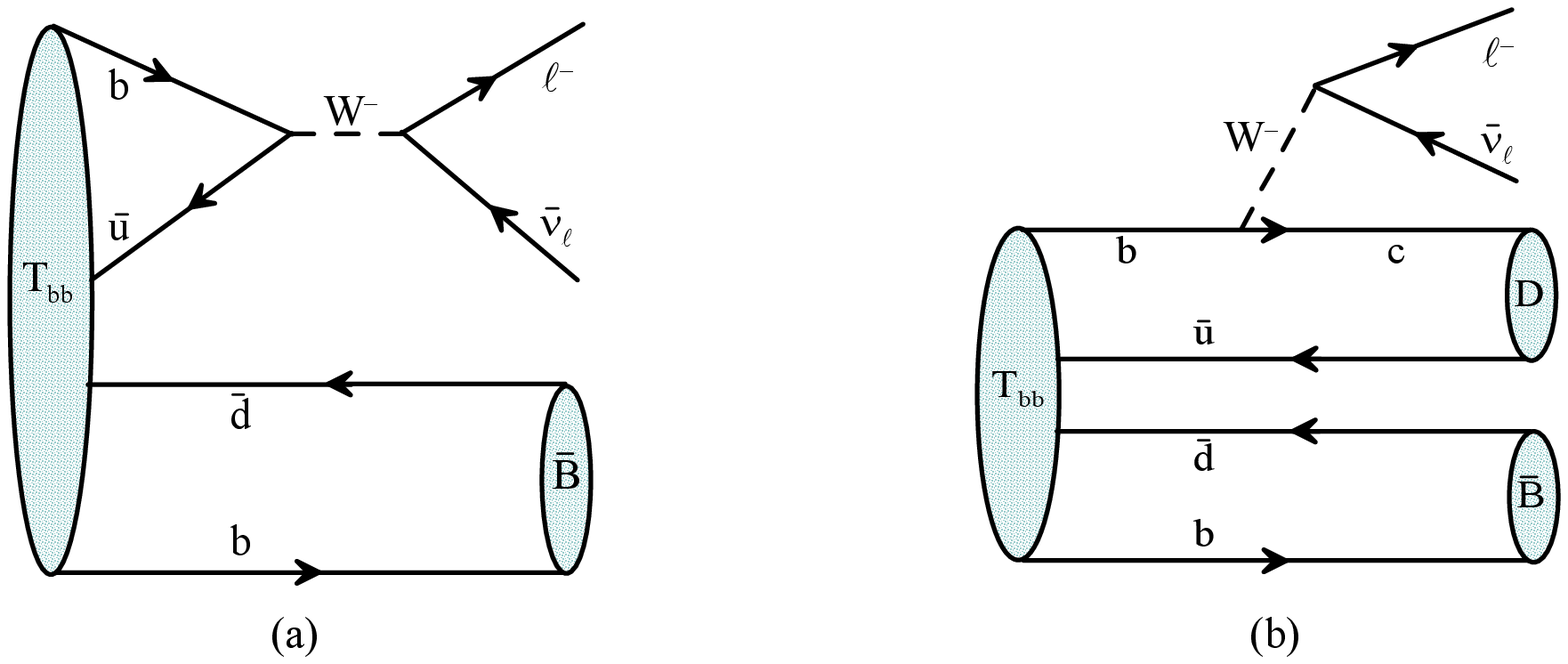}
 \qquad
\includegraphics[width=.47\textwidth]{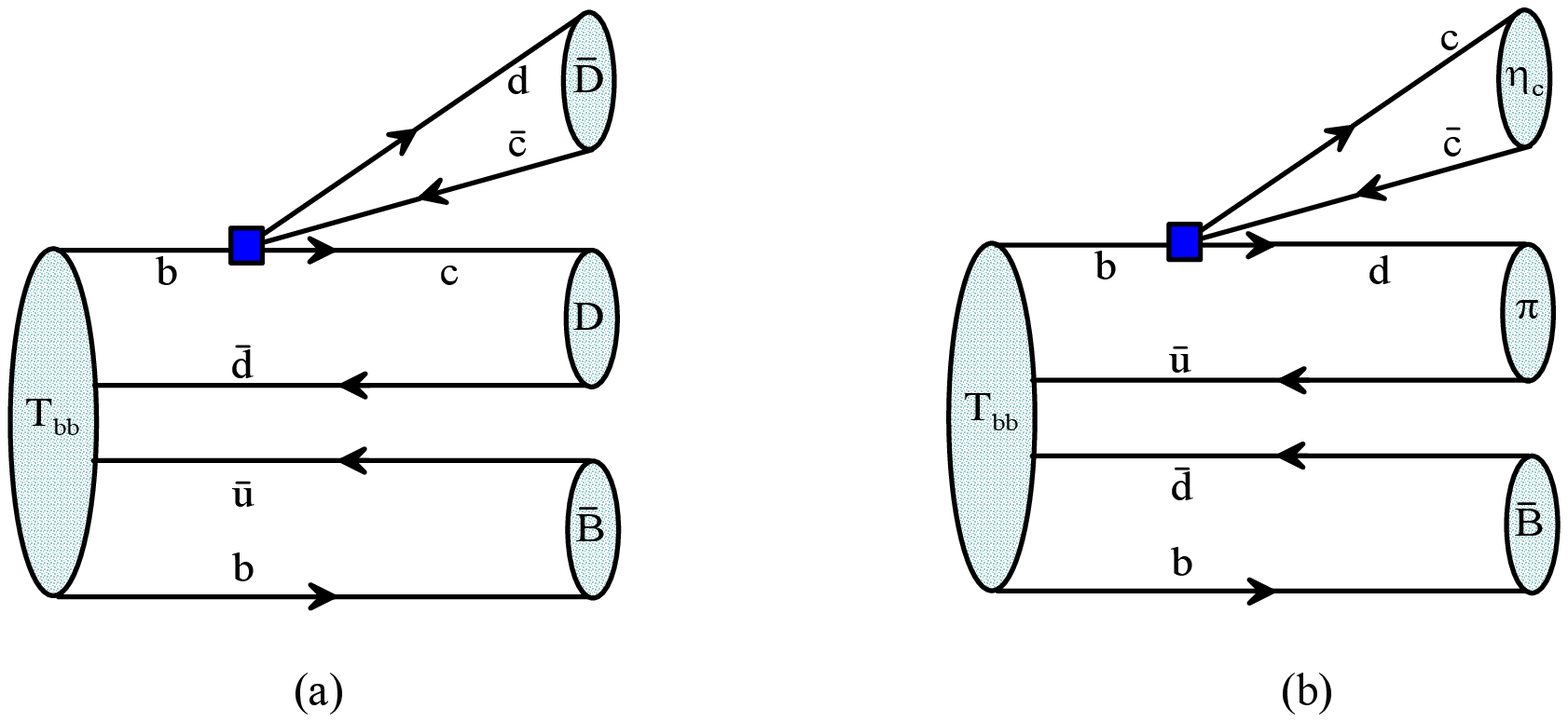}}
 \caption{Representative diagrams for semileptonic (left) and 
nonleptonic (right) decays of the $T_{bb}$ tetraquark.}
 \label{FR_8}
\end{figure}

For a representative state below all possible strong-decay thresholds, for example the
$T_{bb}$ tetraquark discussed in Sect.~\ref{sec2-1}, we have calculated its decay width 
due to all plausible semileptonic 
and nonleptonic decay modes~\cite{Her20}. Some of the corresponding 
processes are illustrated in Fig.~\ref{FR_8}. The hadronic decays are 
calculated within the factorization approximation~\cite{Her06}.
The largest partial widths are found to be of the order of $10^{-15}$ to $10^{-14}\,$GeV. For the 
semileptonic modes, the corresponding decays are of the type 
$\bar B{}^{*} D^{*}\,\ell^-\bar\nu_\ell$, where $\ell=e$ or $\mu$.
Due to the large phase space available in all cases, the differences 
among the widths into the three lepton families are very small.

For semileptonic decays with two mesons in the final state, the processes 
involving a $ b \to c$ vertex are favored compared to those involving a $b \to u$ vertex, 
due to the larger CKM matrix element. In Table~\ref{TR_3} we show
the most favorable channels, the filter being a width larger than 
$10^{9}\,{\rm s}^{-1}=0{.}66 \times 10^{-15}$\,GeV,
for the semileptonic decays with two mesons and a light $\ell=e,\mu$ lepton 
in the final state. Though much smaller, we also give the widths for the
corresponding channels with a final $\tau$ since they could be interesting in
the context of studies of lepton-flavor universality violation.
Besides, due to spin recoupling coefficients, the largest decay widths appear for vector 
mesons in the final state.

For the nonleptonic decays, the largest widths are for the decays of 
the type $\bar B{}^{*} D^{*} D_s{}^{*}$. 
All of them contain a $b \to c$ vertex and a $D^{*}_{s}$ meson in the final
state. See Table~\ref{TR_4}. Once again  vector mesons are favored in the final state. 
Processes with $D_s$ or a light meson final state arising from vacuum have 
decay widths comparable to the corresponding semileptonic decay.
\begin{table}[t]
\caption{Decay widths, in units of $10^{-15}$~GeV, of 
the leading semileptonic modes of $T_{bb}$.}
\begin{center} 
\begin{tabular}{lclc}
\hline
Final state &$\Gamma$ &Final state &$\Gamma$\\
\hline
${B^*}^{-} \, {D^*}^{+} \, \ell^- \, \bar \nu_\ell $ & \multirow{2}{*}{$9{.}02 \pm 0{.}07 $}&
${B^*}^{-} \, {D^*}^{+} \, \tau^- \, \bar \nu_\tau $&\multirow{2}{*}{$1.55\pm0.01 $}\\
$\bar {B^*}^{0} \, {D^*}^{0} \, \ell^- \, \bar \nu_\ell $ && $\bar {B^*}^{0} \, {D^*}^{0} \, \tau^- \, \bar \nu_\tau $\\
${B^*}^{-} \, D^{+} \, \ell^- \, \bar \nu_\ell $   & \multirow{2}{*}{$3{.}59 \pm 0{.}03$}&
${B^*}^{-} \, D^{+} \, \tau^- \, \bar \nu_\tau$     &\multirow{2}{*}{$ 0.727\pm0.005$}\\
$\bar {B^*}^{0} \, D^{0} \, \ell^- \, \bar \nu_\ell $&&  $\bar {B^*}^{0} \, D^{0} \, \tau^- \, \bar \nu_\tau $ \\
$B^{-} \, {D^*}^{+} \, \ell^- \, \bar \nu_\ell $ &\multirow{2}{*}{$4{.}63 \pm 0{.}05$}&
$B^{-} \, {D^*}^{+} \, \tau^- \, \bar \nu_\tau $&\multirow{2}{*}{$ 0.86\pm0.007$} \\
$\bar B^{0} \, {D^*}^{0} \, \ell^- \, \bar \nu_\ell $&&$\bar B^{0} \, {D^*}^{0} \, \tau^- \, \bar \nu_\tau $\\
$B^{-} \,  D^{+} \, l^- \, \bar \nu_l $& \multirow{2}{*}{$1{.}92 \pm 0{.}02$}&
$B^{-} \,  D^{+} \, \tau^- \, \bar \nu_\tau $&\multirow{2}{*}{$ 0.409\pm0.003$}\\
$\bar B^{0} \, D^{0} \,  \ell^- \, \bar \nu_\ell $&&$\bar B^{0} \, D^{0} \,  \tau^- \, \bar \nu_\tau $\\ 
\hline
\end{tabular}
\end{center}
\label{TR_3}
\end{table}
\begin{table}[b]
\caption{Decay widths, in units of $10^{-15}$\,GeV, of the leading 
nonleptonic modes of $T_{bb}$.}
\begin{center}
\begin{tabular}{lclc}
\hline
Final state &$\Gamma$ &Final state &$\Gamma$\\
\hline
${B^*}^{-} \, {D^*}^{+} \, D_s^-$       & \multirow{2}{*}{$4{.}00 \pm 0{.}06$} &
${B^-} \, {D^*}^{+} \, {D_s^*}^- $      & \multirow{2}{*}{$3{.}15 \pm 0{.}05$}\\
$\bar {B^*}^{0} \, {D^*}^{0} \, D_s^- $                         && $\bar {B^0} \, {D^*}^0 \, {D_s^*}^- $  \\  
${B^*}^{-} \, {D^*}^{+} \, {D_s^*}^-$  & \multirow{2}{*}{$6{.}50 \pm 0{.}09$} &
$B^- \, D^+ \, {D_s^*}^- $    & \multirow{2}{*}{$1{.}20 \pm 0{.}02$}\\
$\bar {B^*}^{0} \, {D^*}^{0} \, {D_s^*}^- $ &&          $\bar {B^0} \, D^0 \, {D_s^*}^- $\\
${B^*}^{-} \, D^{+} \, D_s^-$& \multirow{2}{*}{$2{.}57 \pm 0{.}04$} &${B^*}^{-} \, {D^*}^{+} \, \rho^-$  & $3{.}57 \pm 0{.}09$ \\
$\bar {B^*}^{0} \, D^0 \, D_s^- $&& ${B^*}^{-} \, {D^*}^{+} \, \pi^-$& $1{.}28 \pm 0{.}03$ \\
${B^*}^{-} \, D^{+} \, {D_s^*}^- $ & \multirow{2}{*}{$2{.}32 \pm 0{.}03$} &
${B^*}^{-} \, D^+ \, \rho^-$                                    & $1{.}70 \pm 0{.}04$ \\
$\bar {B^*}^{0} \, D^0 \, {D_s^*}^- $ &&${B^*}^{-} \, D^+ \, \pi^-$& $0{.}70 \pm 0{.}02$ \\
$B^- \, {D^*}^{+} \, D_s^-$& \multirow{2}{*}{$2{.}78 \pm 0{.}05$} &$B^- \, {D^*}^{+} \, \rho^-$& $2{.}01 \pm 0{.}05$ \\
$\bar {B^0} \, {D^*}^{0} \, D_s^- $&& $B^- \, {D^*}^{+} \, \pi^-$& $0{.}77 \pm 0{.}03$ \\
\hline
\end{tabular}
\end{center}
\label{TR_4}
\end{table}

Finally, as it has been suggested the possible existence
of an strong-stable isoscalar $T_{bc}$ 
tetraquark with quatum numbers $J^P=0^+$~\cite{Car19,Kar17}, 
we evaluate the decay $T_{bb} (1^+) \to T_{bc}(0^+)\ell^-\nu_{\ell}$ which is $7.5\times 10^{-15}\,$GeV.
The semileptonic decay to the $0^+$ $T_{bc}$ tetraquark is relevant but
it is not found to be dominant in clear disagreement with the result of Ref.~\cite{Aga19}, 
obtained using a QCD three-point sum rule approach.

We also estimated all plausible decay modes such as $\bar B{}^0 e^-\bar\nu_e$, or $B{}^{*-}D^+\pi^-$, etc. 
The total width turns out to be about $87\times 10^{-15}\,$GeV, which gives an 
upper bound for the lifetime of about $7.6\,$ps. 
This lifetime is one order of magnitude larger than the 
simplest guess-by-analogy estimation of $0.3$\,ps of Ref.~\cite{Kar17}.
It is important to note
that a long lifetime for the $T_{bb}$ tetraquark 
can ease its detection through the method of {\it displaced vertex} 
proposed in Ref.~\cite{Ger18}.

\section{Decay width of $Qq\bar Q\bar q$ states}
\label{sec5}
\begin{figure*}[t]
\begin{center}
\includegraphics[trim={1cm 12cm 1cm 1.2cm},clip,width=.6\textwidth]{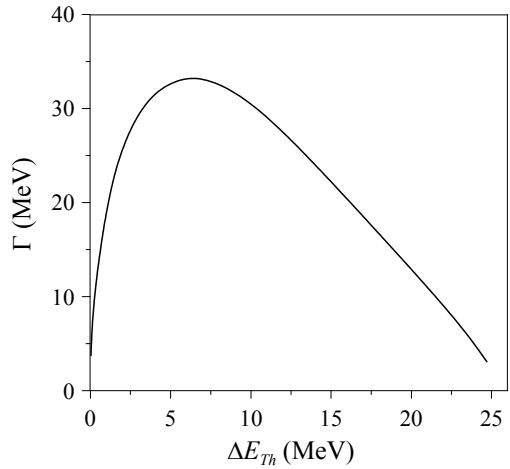}
\end{center}
\caption{Width of the resonance, $\Gamma$, as a function of
the energy difference between its mass and 
the mass of the lower threshold generating the state, $\Delta E_{Th}=M_R-m_1-m_2$.
The upper channel is 25.6~MeV above the lower one.}
\label{FR_9}
\end{figure*}

We have finally addressed the study of the decay width for those cases where a resonance 
is produced between two thresholds, thanks to a coupling between two 
internal configurations within the resonance~\cite{Gar18}. For this purpose, 
we have modeled the system as a coupled-channel problem obeying  the 
non-relativistic Lippmann-Schwinger equation. Channel 1, the lowest in mass, consists of two particles with masses $m_1$ and $m_2$, 
and channel 2, the upper in mass, is made of two particles 
with masses $m_3$ and $m_4$. The Lippmann-Schwinger equation is  written as,
\begin{equation}
t^{ij}(p,p';E)= V^{ij}(p,p')+\sum_{k=1,2} \int_0^\infty {p''}^2
{\rm d}p''
\frac{V^{ik}(p,p'')\,t^{kj}(p'',p';E)}{E-\Delta M \,\, \delta_{2,k}-\dfrac{p''{}^2}{2\,\mu_k}+i\epsilon} 
 \, ,
\label{ER_7} 
\end{equation}
where $i,j=1,2$, $\mu_1=m_1 m_2/(m_1+m_2)$ and 
$\mu_2=m_3 m_4/(m_3+m_4)$ are the reduced masses of channels 1 and 2,
and $\Delta M=m_3+m_4-m_1-m_2$ with $m_3+m_4 > m_1+m_2$.
The interaction kernels in momentum space are given by,
\begin{equation}
V^{ij}(p,p')=\frac{2}{\pi}\int_0^\infty r^2dr\; j_0(pr)V^{ij}(r)j_0(p'r)\,,
\label{ER_8} 
\end{equation}
where the two-body potentials, which are the inputs of the modeling, 
consist of an attractive and a repulsive Yukawa term, i.e.,
\begin{equation}
V^{ij}(r)=-A\frac{e^{-\mu_Ar}}{r}+B\frac{e^{-\mu_Br}}{r}\,.
\label{ER_9} 
\end{equation}
We have considered scenarios where a resonance exists at an 
energy $E=E_R$, such that the phase shift $\delta(E_R)=90^\circ$,
for energies between the thresholds of channels 1 
and 2, i.e., $0 < E_R < \Delta M$. The mass of the 
resonance is given by $M_R=E_R + m_1 +m_2$, and its
width is calculated using the Breit-Wigner formula as~\cite{Bre36,Cec08,Cec13},
\begin{equation}
\Gamma (E) =\lim\limits_{E \to E_R}\, \frac{2(E_R-E)}{\text{cotg}[\delta(E)]} \, .
\label{ER_10} 
\end{equation}
By varying the parameters in Table~\ref{TR_5}, one can control the existence of 
a bound state or a resonance and its relative position with respect to the 
thresholds.
We choose as starting point the set of parameters given 
in Table~\ref{TR_5}. 
\begin{table}[b]
\caption{Parameters of the interaction as given in Eq.~\eqref{ER_9}.
$A$ and $B$ are in MeV fm, while $\mu_A$ and $\mu_B$ are 
in ${\rm fm}^{-1}$. $m_1=m_2=1115.7$ MeV/c$^2$, $m_3=938.8$ MeV/c$^2$, 
and $m_4=1318.2$ MeV/c$^2$.} 
\begin{center}
\begin{tabular}{ccccc} 
\hline
Channel &  $A$    & $\mu_A$ & $B$   & $\mu_B$  \\ \hline
$1\leftrightarrow 1$ &  $100$  & $2.68$  & $667$ & $5.81$ \\ 
$2\leftrightarrow 2$ &  $680$  & $4.56$  & $642$ & $6.73$ \\ 
$1\leftrightarrow 2$ &  $200$  & $1.77$  & $195$ & $3.33$ \\ \hline
\end{tabular}
\end{center}
\label{TR_5} 
\end{table}
They are adjusted such that in a single-channel calculation, the upper channel (channel 2)
has a bound state just at threshold, while in a coupled-channel 
calculation, the full system has a bound state just at the lower 
threshold.
If one increases the magnitude of the repulsive term  in the lower 
channel, $B(1\leftrightarrow 1)$ in Table~\ref{TR_5}, the bound state 
of the coupled-channel system moves up and 
actually becomes a resonance into the continuum. One can study the behavior
of its width when its mass evolves from the lower threshold, channel 1, 
to the upper one, channel 2. The result is shown in Fig.~\ref{FR_9}. 

The width of the resonance starts increasing quickly when
getting away from the lower threshold, but at about a third of the way
towards the upper channel, the width starts to decrease
although the phase space for the decay to channel 1, where the resonance is
observed, still increases\footnote{Although the Breit-Wigner 
formula is not very accurate close to threshold; however, we have explicitly 
checked by analytic continuation of the S-matrix on the second Riemann sheet 
that at low energy the width follows the expected $\Gamma \sim E^{1/2}$ 
behavior, the one shown by Fig.~\ref{FR_9}.}.
It is important to note that the strength of the coupling between the
two thresholds has not been modified. When the resonance
approaches the upper threshold, it becomes narrow  and seemingly ignores 
the existence of the lower threshold. The wave function of the $(m_3,m_4)$ bound 
state of vanishing energy has, indeed, little overlap with the $(m_1,m_2)$ configuration. 
The same trend is obtained for different strengths of the coupling interaction 
in Table~\ref{TR_5} or varying the mass difference between the two 
thresholds~\cite{Gar18}. Hence, in this region, the dynamics is dominated by the attraction 
in the upper channel and the second channel is mainly a tool for the detection.  
This mechanism is somewhat related to the 'synchronization of resonances' proposed 
by D.~Bugg~\cite{Bug08}.
\begin{figure*}[t]
\begin{center}
\vspace*{-0.5cm}
\resizebox{8.cm}{12.cm}{\includegraphics{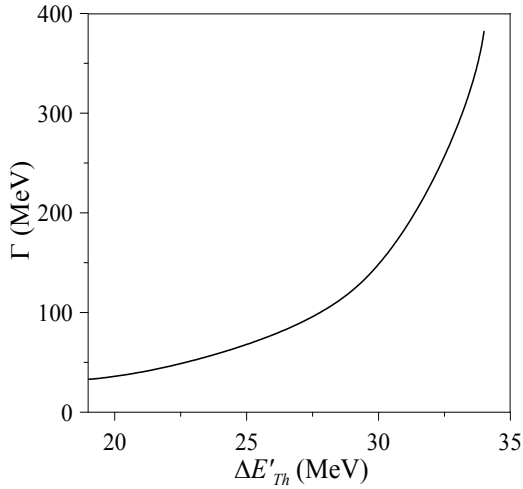}}
\vspace*{-5.0cm}
\caption{Width of the resonance, $\Gamma$, as a function of
the energy difference between its mass and the mass of the upper threshold
generating the state, $\Delta E^\prime_{Th}=m_3+m_4-M_R$, for a fixed 
energy with respect to the lower threshold, $\Delta E_{Th} = M_R-m_1-m_2=$ 6.5~MeV.}
\label{FR_10}
\end{center}
\end{figure*}

The mechanism we have discussed above could help to understand the narrow width of 
several of the hidden heavy-flavor resonances with a large phase
space in a decay channel that have been recently reported 
in both the meson and baryon sectors,
whose hypothetical internal structure would  
allow them to split into different 
subsystems~\cite{Che16,Bri16,Ric16,Leb17,Ali17,Esp17,Liu19}. This would apply, for example,
to the $B \bar B^*$ state of Fig.~\ref{FR_5}(b). 
The situation resembles a Feshbach resonance, where the open channel is represented
by the $\Upsilon \,\omega$ state that would get trapped in a molecular
state supported by the closed channel potential $B \bar B^*$~\cite{Bra04,Pil14}.
An unexpected behavior of the width 
of the resonance may be indicating an important contribution of 
coupled-channel dynamics and the knowledge of the decay width in
a particular channel would hint to the upper threshold contributing to the
formation of the resonance. This has been illustrated in Fig.~\ref{FR_10}, where we 
have calculated the width of the resonance for a fixed value of its mass
with respect to the lower threshold, $\Delta E_{Th} = M_R-m_1-m_2=$ 6.5 MeV, but increasing
the distance with respect to the upper threshold, $\Delta E^\prime_{Th} = m_3+m_4-M_R$. For this purpose,
we have diminished the mass of the lower channel in steps of 5 MeV, thus increasing
the distance between thresholds, $m_3+m_4-m_1-m_2$, and we have increased
$A(1\leftrightarrow 1)$ in Table~\ref{TR_5} in such a way that $\Delta E_{Th} =M_R-m_1-m_2=$ 6.5~MeV 
remains constant. The result is striking, being the phase space fixed for the detection channel,
the width increases when the upper threshold moves away. Thus the width provides
also with basic information about the coupled channels that
may contribute to the formation of a resonance. The observation of a small 
width in a low-lying channel hints 
to a dominant contribution of some upper channel to the formation of the resonance.
Thus, although the exact shape of the dependence of the
width on its position with respect to the detection channel would depend 
on the specific dynamics of the coupled-channel system,
the gross features reflected here might be a relevant and basic 
hint to explore the nature of some of the exotic states.

\begin{acknowledgements}
The authors are deeply indebted to their long-term collaborators 
T.~F.~Caram\'es, E.~Hern\'andez, J.~-M.~Richard 
and J.~Vijande that have participated in some of the issues reviewed in this work. 
This work has been partially funded by COFAA-IPN (M\'exico)
and by Ministerio de Econom\'\i a, Industria y Competitividad 
and EU FEDER under Contracts No. FPA2016-77177 and RED2018-102572-T.
\end{acknowledgements}

\end{document}